\documentstyle[preprint,prb,aps]{revtex}
\begin{document}
\draft
\title{A Monte Carlo
study of temperature-programmed desorption spectra with attractive
lateral interactions.}
\author{A.P.J. Jansen\cite{curadd}}
\address{Laboratory of Inorganic Chemistry and Catalysis\\
Eindhoven University of Technology\\
P.O. Box 513\\
5600 MB Eindhoven\\
The Netherlands\\
E-mail address: tgtatj@chem.tue.nl}
\date{April 26, 1995}
\maketitle
\begin{abstract}
We present results of a Monte Carlo study of temperature-programmed
desorption in a model
system with attractive lateral interactions.
It is shown that even for weak interactions there are large shifts
of the peak maximum temperatures with initial coverage.
The system has a transition temperature below which the desorption
has a negative order.
An analytical expression for this temperature is derived.
The relation between the model and real systems is discussed.
\end{abstract}
\section{Introduction.}

Many studies have been published on Monte Carlo (MC) simulation of
de\-sorption from single-crystal
surfaces,\cite{sal87,gup88,lom88,sal89,gup89,lom89,lom91,men94}
but little attention has been given to the effect of
attractive lateral interactions.
There may be three reasons for this.
First, temperature-programmed desorption (TPD) spectra with
attractive lateral interactions appear less
interesting than TPD spectra with repulsive lateral interactions,
which can show multiple peaks even with just one adsorbate.
Second, repulsive lateral interactions are more common than
attractive lateral interactions.
This can be understood using the bond-order conservation
principle.\cite{shu86}
Nearest neighbor interactions are often repulsive, because nearest
neighbors bond to the same substrate atom(s).
Multiple bonds of a substrate atom are generally weaker than when a
substrate atom bonds just to one adsorbate, so it's energetically
more favorable for adsorbates to stay apart.
Only when the direct adsorbate-adsorbate interaction is strong with
respect to the substrate-adsorbate interaction then one will find
attractive lateral interactions.
This is the case for physisorbed adsorbates.
Third, a good quantitative MC method is necessary to understand the
effects of attractive lateral interactions.
In particular, it is necessary to have an MC method with the correct
real time, instead of MC time, dependence.
Such methods have only recently been developed.\cite{men94,jan95}
It is known that attractive lateral interactions cause peaks in TPD
spectra to shift to higher temperature and to become
smaller.\cite{sal87,gup88,gup89}
In this paper we will concentrate on shifts of peaks with initial
coverage.
\section{Computational details.}

We have used MC modeling of the reactions based on the master
equation
\begin{equation}
  {dP_\alpha\over dt}
  =\sum_\beta
  \big[W_{\alpha\beta}P_\beta
  -W_{\beta\alpha}P_\alpha\big],
\end{equation}
where $P_\alpha$ is the probability of having the adlayer in
configuration $\alpha$.
The $W$'s are transition probabilities per unit time corresponding to
various reactions.
The master equation is solved using the first-reaction method in
which a time for each reaction to occur is generated from an
appropriate probability distribution, and then the adlayer is
modified according to the first reaction after which the procedure
is repeated.\cite{gil76}
Details of this method can be found elsewhere.\cite{jan95}
The transition probabilities per unit time are given by
\begin{equation}
  W_{\alpha\beta}=\nu_{\alpha\beta}
  \exp\left[-
  {E_{\alpha\beta}\over k_BT}\right],
\end{equation}
where $E_{\alpha\beta}$ and $\nu_{\alpha\beta}$ are the activation
energy and prefactor, respectively, of the reaction
$\beta\rightarrow\alpha$ that transforms
configuration $\beta$ into configuration $\alpha$.
The $W$'s may be time-dependent through the temperature.
For the simulation of TPD spectra
\begin{equation}
  T=T_0+Bt
\end{equation}
holds, where $T_0$ is the
temperature at the begin of the simulation, $t$ is time, and $B$ is
the (constant) heating rate.

For all the desorption reactions we have used the same prefactor
$\nu_{\alpha\beta}=\nu_0$.
The activation energy is given by
\begin{equation}
  E_{\alpha\beta}=E_0(1+\gamma N_{\alpha\beta}),
  \label{eA}
\end{equation}
where $E_0$ is the activation energy for an isolated adsorbate,
$N_{\alpha\beta}$ is the number of (nearest) neighbors of the
desorbing adsorbate in the reaction $\beta\rightarrow\alpha$, and
$\gamma$ is a positive dimensionless parameters determining the
strength of the interactions between the adsorbates.
The linear dependence of the activation energy on the number of
neighbors seems to be appropriate for the systems that are best
known for attractive adsorbate-adsorbate interaction, namely noble
gas atoms, but we will also look at other dependencies on the number
of neighbors later on.

Apart from desorption we also have diffusion.
This is modeled by jumps of adsorbates to neighboring sites.
The diffusion should be fast with respect to desorption, because we
want a thermally equilibrated adlayer before each desorption.
We found that for values $\gamma<0.1$ a prefactor for diffusion of
$\nu_{\alpha\beta}=20\nu_0$ was sufficient.
We used the same expression~(\ref{eA}) for the activation energy as
for the desorption.
The number of neighbors in that expression refers to the situation
prior to the jump of the adsorbate.

Most MC simulations were done on a square lattice of dimensions
$100\times 100$.
This effectively models an infinite lattice.
Although we didn't do extensive simulations of lattices with different numbers
of neighboring sites $Z$, we will present an analysis that holds
for any value of $Z$.
All simulations have been done with our code PIZZAZZ.\cite{jan94}
\section{Results and discussion.}

The MC simulations that have just been described depend on five
parameters; $\nu_0$, $E_0$, $\gamma$, $B$, and $T_0$.
The last one is not very relevant; as long as it is chosen as a
low temperature
where the desorption is negligible the TPD spectra will be the same.
We can remove two more parameters by scaling the time and the
temperature.
Instead of time we use $\nu_0t$, and instead of temperature we use
$k_BT/E_0$.
The only remaining parameters are then $\gamma$ and $k_BB/\nu_0E_0$,
the scaled version of $B$.
The former is, of course, the more relevant one.
We will also present results for desorption at constant temperature.
These depend on just one parameter; $k_BT/E_0\gamma$.

Fig.~\ref{f1} shows the TPD spectrum for $\gamma=0.05$ and
$k_BB/\nu_0E_0=10^{-16}$.
(This is roughly the TPD spectrum one would get with
$\nu_0=10^{13}$s${}^{-1}$, $E_0/k_B=3000\,$K, $E_0\gamma/k_B=150\,$K,
and $B=3\,$K.s${}^{-1}$, values very similar to the Xe/Pt(111)
system.\cite{jan92})
Although $\gamma$ is rather small, we see a substantial shift of
the peak maximum with decreasing initial coverage.
Lower coverages may show larger desorption rates.
This implies that the order of the desorption is negative,
which has been observed in Xe/Pt(111) experimental spectra
before.\cite{jan92,sid90}
Smaller values of the parameter $\gamma$ show smaller distances
between the peak
maxima, but not until $\gamma$ is clearly smaller than 0.01
the curves of different initial coverages stop crossing.
(We will derive a better estimate below.)
Making $\gamma$ larger spreads out the peak as far as we could check.
However, for $\gamma$ larger than about 0.1 we ran into convergence
problems with the grid size and the diffusion rate.
For really large values of $\gamma$ one expects island formation.
Islands are one explanation for zero-order desorption,\cite{yat85}
so the trend towards more negative order should reverse for
some $\gamma$.

As the interactions are attractive between the adsorbates, there is a
order-disorder phase transition.
Below the phase-transition temperature islands just mentioned
are formed.
This phase-transition temperature is about half $E_0\gamma/k_B$.
The negative-order TPD spectra have peaks well
above this temperature, however.
This implies that a model, which assumes that the
adsorbates are randomly distributed over the substrate, might explain
the peak shifts.
To check this it seems more convenient to look at desorption at a
constant temperature,
as this reduces the relevant parameters to just one; $k_BT/E_0\gamma$.
The results are shown in Fig.~\ref{f2}.
The model and the MC results agree very well except for the
relatively low temperatures at intermediate coverages.
At those temperatures there is still some short-range order.
The adsorbates tend to have somewhat more neighbors than in a random
distribution so that the desorption is smaller than in the model.
The model and the MC simulation show that for desorption at low
constant temperature starting with a high coverage the desorption
rate increases, reaches a maximum, and then decreases.
The initial increase implies a negative order.
At higher temperatures the order is positive for all coverages, and
so we have another transition temperature.
This temperature can be defined as the temperature where the order is
zero at $\theta=1$.

We can derive this temperature from the model, or more generally, by
assuming that for $\theta\approx 1$ there are only adsorbates which
still have all their neighbors or have only lost one.
With $N_\alpha$ the number of adsorbates in configuration $\alpha$
the master equation gives us the general result
\begin{eqnarray}
  {d\langle N\rangle\over dt}
  &&\equiv{d\over dt}\sum_\alpha P_\alpha N_\alpha\nonumber\\
  &&=\sum_{\alpha\beta}W_{\alpha\beta}
   P_\beta(N_\alpha-N_\beta)\nonumber\\
  &&=-\sum_\beta P_\beta\sum_{n=0}^ZW^{(n)}N_\beta^{(n)},
\end{eqnarray}
where $N_\beta^{(n)}$ is the number of adsorbates in configuration
$\beta$ with $n$ neighbors, and $W^{(n)}$ is the
transition probability per unit time for such adsorbates.
Dividing by the number of sites this gives us for
the desorption rate $r$ at $\theta\approx 1$ the expression
\begin{eqnarray}
  r&&\equiv
  -{d\theta\over dt}\\
  &&=\big[1-(1+Z)(1-\theta)\big]W^{(Z)}+Z(1-\theta)W^{(Z-1)},\nonumber
\end{eqnarray}
where $Z$ is the number of neighboring sites.
The coefficients of $W^{(n)}$ are the fractions of all adsorbates
with $n$ neighbors.
{}From this we can find when the negative-order desorption disappears
by setting $dr/dt=-d^2\theta/dt^2=0$ at $\theta=1$, which gives us
\begin{equation}
  {W^{(Z)}\over W^{(Z-1)}}
  ={Z\over Z+1}.
  \label{eB}
\end{equation}
As $W^{(n)}\propto\exp(-nE_0\gamma/k_BT)$, this leads to
\begin{equation}
  k_BT={E_0\gamma\over\ln(Z+1)-\ln Z}
\end{equation}
for the transition temperature.
For $Z=4$ we have $k_BT\approx 4.48\,E_0\gamma$, which is indeed far
above the order-disorder phase-transition temperature.
It's also instructive to compute the order $x$ at $\theta=1$.
We find that it equals for desorption at a constant temperature
\begin{eqnarray}
  x\equiv
  {\theta\over r}\left(\partial r\over\partial\theta\right)_T
  &&=\theta{d^2\theta\over dt^2}\left/
  \left({d\theta\over dt}\right)^2\right.\nonumber\\
  &&=1-Z\left[e^{E_0\gamma/k_BT}-1\right].
\end{eqnarray}
For the values used in Fig.~\ref{f1} this becomes $x=-9.96$ at the
peak maximum temperature of $k_BT/E_0=0.0379$.
Whether the negative order really shows up in a TPD spectrum depends
on the positions of the peaks.
For normal heating rates the peak maximum is at about
$k_BT/E_0\approx 1/30$.
This value should be lower than the temperature at which the negative
order disappears.
Hence for $Z=4$ we find negative orders in TPD spectra if $\gamma
>0.00744$.

We would finally like to remark on the relevance of the results just
presented to real systems.
As has already been mentioned negative orders in TPD have been seen
for Xe/Pt(111).\cite{jan92,sid90}
As the effect seems to be large, it should be easy to find
experimentally.
However, this depends crucially on the dependence of the activation
energy for the desorption on the number of neighbors.
Above we have assumed that the activation energy was additive.
If we assume a dependence $E_0(1+\gamma\sqrt{N_{\alpha\beta}})$ then
we find from Eq.~(\ref{eB})
\begin{equation}
  k_BT=E_0\gamma
  {\sqrt{Z}-\sqrt{Z-1}\over\ln(Z+1)-\ln Z}
\end{equation}
for the transition temperature.
For $Z=4$ this gives us $k_BT=1.20\,E_0\gamma$, and with the peak
maximum at $k_BT/E_0\approx 1/30$ we must have $\gamma>0.0278$
in order to get negative orders in TPD spectra.
We see that the effects are much weaker than for additive
interactions.
As attractive interactions are primarily found for physisorption,
the additive activation energy may seem to be a good approximation.
However, substrate-mediated effects may give large non-additive
contributions.\cite{jan92}

Non-additivity need not be the only factor obscuring negative orders
as we can demonstrate for Xe/Pt(111).
Activation energies for this system as a function of the number of
neighbors have
been determined using molecular dynamics (MD).\cite{jan92}
They are well approximated by $E_0/k_B=2851\,$K and $\gamma=0.031$.
As $Z=6$, negative-order desorption should disappear only at
$575\,$K, whereas peak maxima are found somewhat below $110\,$K.
Large peak shifts as a function of initial coverage are expected.
Indeed, if we use the same activation energies for the diffusion,
we find spectra very similar to Fig.~\ref{f1}.
However, such large shifts have neither been found
experimentally,\cite{sid90} nor with MD.\cite{jan92}
The reason for this is that the lateral interactions are larger
than the activation energies suggest.
A value of $\gamma=0.049$ is more appropriate for diffusion.
This is found using the effective Xe--Xe potential of the MD
study.
If we use this value for the diffusion we obtain the spectra in
Fig.~\ref{f3}.

We see that the shifts are small.
There is a reasonable agreement between Fig.~\ref{f3} and the
zero-order-like spectra of the MD study.
Figure~\ref{f3} still shows slight negative order, and there are no
extra peaks or shoulders.
We think that these latter features in the MD study are probably
artifacts caused by fits and extrapolations.
The same holds for the oscillations in the order and the prefactor
(Fig.~7 of the MD study\cite{jan92}), which are particularly
sensitive to numerical errors.

It might seem contradictory that attractive lateral interactions,
that in our model with randomly distributed adsorbates cause
large peak shifts, reduce these shifts in Xe/Pt(111).
The reason for this is that in the model the lateral interactions
only effect the desorption, whereas the effect in Xe/Pt(111) is
dominantly on the structure of the adlayer.
For the model the probability that an adsorbate has
$n$ neighbors equals $\left({Z\atop n}\right)\theta^n
(1-\theta)^{Z-n}$ (note that this does not depend explicitly
on the temperature).
For Xe/Pt(111) the fraction of Xe atoms with many neighbors
is larger than according to this expression, as desorption takes
place not far above the order-disorder transition.
The probability of a Xe atom having $n$ neighbors and the short-range
order are similar to what was found in the MD study.
Figure~\ref{f2} already showed the consequence.
The curves of the model are always above the results of the
simulations.
At low temperature this means, that starting at high coverage, the
desorption rate in the model increases more than in the simulations,
which implies a more negative order, and hence larger peak shifts.
\section{Conclusions.}

TPD spectra of systems with attractive lateral interactions may show
large shifts of the peak maxima as a function of initial
coverages.
These shifts indicate negative-order desorption.
A simple model with random occupation of sites can explain them
qualitatively.
The same model can also be used to derive an exact analytical expression
for a transition temperature above which negative-order desorption
is not possible.
This transition temperature may be higher than the order-disorder
phase-transition temperature by about an order of magnitude.
Non-additivity and the structure of the adlayer may decrease the peak
shifts.
Therefore more experiments are necessary to see if the negative-order
desorption that occurs in the model system described above also
occurs in real systems.
\begin{figure}
\caption{TPD spectra showing the desorption rate $-d\theta/d(\nu_0t)$
as a function of the temperature $k_BT/E_0$ for $\gamma=0.05$ and
heating rate $k_BB/\nu_0E_0=10^{-16}$.
The dots are results directly from the MC simulations, and the lines
connect MC results with the same initial coverage.
The five curves are from left to right for initial coverage
$0.262$, $0.463$, $0.644$, $0.822$, and $1.000$.
\label{f1}}
\end{figure}
\begin{figure}
\caption{Scaled desorption rate $-\exp(E_0/k_BT)\,d\theta/d(\nu_0t)$
as a function of coverage $\theta$.
The symbols are results directly from the MC simulations, and the
lines are from the model with random distribution of adsorbates.
Results are shown for five temperatures from top to bottom
$k_BT/E_0\gamma=6$, $4.48$, $3$, $1.6$, and $0.8$.
The value of $k_BT/E_0\gamma=4.48$ corresponds to the transition
temperature above which negative-order desorption is not possible.
\label{f2}}
\end{figure}
\begin{figure}
\caption{TPD spectra showing the desorption rate $-d\theta/dt$ (in
number of Xe atoms per second per site)
as a function of the temperature (in K) for heating rate
$B=1\,$K$\cdot$s${}^{-1}$.
The dots are results directly from the MC simulations, and the lines
connect MC results with the same initial coverage.
The five curves are from left to right for initial coverage
$0.598$, $0.698$, $0.806$, $0.906$, and $1.000$.
\label{f3}}
\end{figure}

\begin{references}
\bibitem{jan94} PIZZAZZ is a general-purpose program, written in C++,
for simulating reactions on surfaces that can be represented by
regular grids.
\end{references}
\end{document}